\documentclass[12pt,preprint]{aastex}

\begin{document}
\title{On the Metallicities of {\it Kepler} Stars}

\author{
Subo Dong\altaffilmark{1},
Zheng Zheng\altaffilmark{2},
Zhaohuan Zhu\altaffilmark{3,4},
P. De Cat\altaffilmark{5},
J.N. Fu\altaffilmark{6},
X.H. Yang\altaffilmark{6,7},
Haotong Zhang\altaffilmark{7},
Ge Jin\altaffilmark{8},
Yong Zhang\altaffilmark{8}
}
\altaffiltext{1}{Kavli Institute for Astronomy and Astrophysics, Peking University, Yi He Yuan Road 5, Hai Dian District, Beijing, 100871, China}
\altaffiltext{2}{Department of Physics and Astronomy, University of Utah, Salt Lake City, UT 84112, USA}
\altaffiltext{3}{Department of Astrophysical Sciences, Princeton
University, Princeton, NJ, 08544}
\altaffiltext{4}{Hubble Fellow}
\altaffiltext{5}{Royal observatory of Belgium, Ringlaan 3, B-1180 Brussel, Belgium}
\altaffiltext{6}{Department of Astronomy, Beijing Normal University, 19 Avenue Xinjiekouwai, Beijing 100875, China}
\altaffiltext{7}{National Astronomical Observatories, Chinese Academy of Sciences, Beijing 100012, China}
\altaffiltext{8}{University of Science and Technology of China, Hefei 230026, China}
\altaffiltext{8}{Nanjing Institute of Astronomical Optics \& Technology, National Astronomical Observatories, Chinese Academy of Sciences, Nanjing 210042, China}

\begin{abstract} 
{We use 12000 stars from Large Sky Area Multi-Object Fiber Spectroscopic Telescope (LAMOST) 
spectroscopic data to show that the 
metallicities of {\it Kepler} field
stars as given in the Kepler Input Catalog (KIC) systematically
underestimate both the true metallicity and the dynamic range of
the {\it Kepler} sample.  Specifically,  
to the first order approximation, we find $$
{\rm [Fe/H]}_{\rm KIC} = -0.20 + 0.43{\rm [Fe/H]}_{\rm LAMOST},
$$
with a scatter of $\sim 0.25$ dex, due almost entirely to errors in KIC.
This relation is most secure for $-0.3<{\rm [Fe/H]}_{\rm LAMOST}<+0.4$
where we have $>200$ comparison stars per 0.1 dex bin and good consistency is shown between metallicities determined
by LAMOST and high-resolution spectra. It remains
approximately valid in a slightly broader range. When the relation
is inverted, the error in true metallicity as derived from KIC is
(0.25 dex)/0.43 $\sim 0.6$ dex.  We thereby quantitatively confirm the cautionary note by \citet{KIC} that KIC estimates of [Fe/H] should not be used by ``anyone with a particular interest in stellar metallicities''}.
Fortunately, many more LAMOST spectroscopic metallicities will be 
available in the near future.
\end{abstract}

\keywords{planetary systems -- stars: abundances}

\section{{Introduction}
\label{sec:intro}}

Of the several thousand planetary candidates found by {\it Kepler},
only a few hundred have high-resolution spectra of their hosts.
The number of ``control sample'' stars (without known planets)
with such spectra is much smaller.
Hence, large-sample statistical studies generally must rely on the
Kepler Input Catalog (KIC, \citealt{KIC}). It is well known that the KIC was not designed for this purpose, and KIC metallicities are known to be particularly problematic. {\citet{KIC} cautioned that ``anyone with a particular interest in stellar metallicities should not use the KIC for their estimates of log(Z).''  Using stellar parameters determined from
34 high-resolution spectra of {\it Kepler} target stars, they found that 
KIC metallicities were $\sim 0.17$ dex smaller and there were  
indications of significant systematics.} But the faintness of {\it Kepler} stars has meant that high-resolution spectra are expensive in telescope
resources. 

An alternate approach is to obtain medium resolution spectra, which
are generally adequate for estimating basic stellar parameters, i.e.,
effective temperature $T_{\rm eff}$, gravity $\log g$, and metallicity
[Fe/H].  Medium resolution spectrographs have the advantage that
they can be easily multiplexed.
For example, the Sloan Digital Sky Survey (SDSS) in its various
incarnations has characterized of order $6\times10^5$ stars using an
$R\sim 2000$ multi-object
optical spectrograph \citep{dr8,dr10}.
Unfortunately, SDSS did not target the {\it Kepler}
field with its optical spectrograph, although SDSS-III has begun
observing brighter {\it Kepler} stars with its high-resolution APOGEE
infrared multi-object spectrograph.

Because of the high science value of planetary hosts, exceptional
efforts have nevertheless been made to obtain spectra.
 \citet{buchhave12} obtained high-resolution spectra for 152 hosts,
  and \citet{everett13} obtained $R\sim 3000$ optical spectra for 268 hosts.
However, because these samples are still relatively small, and more
importantly because the stellar parameters of the underlying population
(with and without planets) is poorly characterized, it is difficult
to do statistical studies of planet frequency as a function of stellar
parameters.

Therefore, large statistical studies have been compelled to make
use of KIC parameters.  For example, \citet{wang13} used KIC
metallicities as a proxy for spectroscopic metallicities 
to estimate the relative
planet frequency for high-vs-low metallicity stars in several
planet-radius bins. Other statistical works using KIC metallicities
include \citet{laughlin}, \citet{dodson} and \citet{dawson}.

The Large Sky Area Multi-Object Fiber Spectroscopic Telescope (LAMOST, 
a.k.a. Goushoujing telescope) is an ideal instrument to explore 
 the $\sim 115\,{\rm deg}^2$
Kepler field with spectroscopy. LAMOST is a Schmidt telescope 
 with a $\sim 4$m effective aperture and 4000 fibers that can be 
deployed a $5^\circ$ diameter field of view.
Here we use data from {Data Release 1 (DR1) and Data Release 2 (DR2)} from LAMOST 
\citep{lamost1, lamost2, lamost_pilot} with $R\sim 1800$ to evaluate the
relation between KIC metallicities and those determined from spectroscopy.  
We show that,
in the mean, there is fairly tight relation, but that the slope
of this relation is quite shallow (0.43).  Thus although the
scatter of KIC metallicities around the true ones is modest (0.25 dex),
if one is compelled to infer the true metallicity from the KIC value,
the error is much larger: 0.25/0.43 $\sim 0.58$ dex.  {Hence we quantitatively confirmed the warning issued by \citet{KIC} that KIC metallicities
must be used with extreme caution.}

{LAMOST DR1 and DR2 reports stellar parameters for $\sim 17000$ {\it Kepler} stars with no preference
for known planet hosts as part of the ``LAMOST-{\it Kepler} project'' 
to observe all target stars in the {\it Kepler} field \citep{lamostkep}.}
The LAMOST samples should eventually enable solid statistical investigations
that are able to accurately characterize both the ``numerators'' (targets hosting planets) and
the ``denominators'' of various subsamples.  
We ourselves are working on analyses regarding 
dependence of planet frequency on metallicities and various other host properties.   
However, our purpose here is to apply DR1 and DR2 Season 1 to a much
more limited question: {quantifying the systematics of KIC metallicities.}

\section{{LAMOST {\it Kepler} Sample}
\label{sec:lamost}}

{We query the LAMOST DR1 and DR2 AFGK-type stars catalog \footnote{http://www.lamost.org/public/survey/datarelease} for {\it Kepler} stars, but not those that were specifically
targeted because they had planets. We find 16959 stars with KIC identifications, of which 317, or 
about 1.9\%, host planetary candidates. This is statistically indistinguishable 
from the  {\it Kepler} catalog as a whole, which has 2716 candidate hosts out 
of $\sim 150,000$ stars, or 1.8\%.
We eliminate those with LAMOST $\log g<3.5$ in order
to focus on dwarf stars.  And we also eliminate stars that lack KIC metallicities.  
This leaves a sample of 12400 stars.  Of these, 64 stars
lie outside the range $-0.85<{\rm [Fe/H]}<+0.65$.  At these extremes, 
there are fewer than 20 stars per 0.1 dex bin, which would lead to
poor statistical precision.  We therefore also eliminate these 64 stars.}

The $T_{\rm eff}$, $\log g$ and [Fe/H] in the catalog are determined by the LAMOST Stellar 
Parameter pipeline. {This pipeline has been built upon the algorithm in \citet{lamost_pipeline} analysing the commissioning LAMOST data, but it has been significantly improved since then, 
in particular taking considerable care in handling problems 
associated with relative flux calibration of the LAMOST spectra, which was found to be a main source of systematics shown in the commissioning data analysed by \citet{lamost_pipeline} before (Private Communication with Ali Luo, 2014).} In Section~\ref{sec:high}, we find that the  
[Fe/H] measurements from the LAMOST DR1 and DR2 catalog have a high degree of 
consistency with those determined from high-resolution spectroscopy in 
the literature.

\section{Comparison of LAMOST to KIC Metallicities
\label{sec:kic}}

\begin{figure}
\plotone{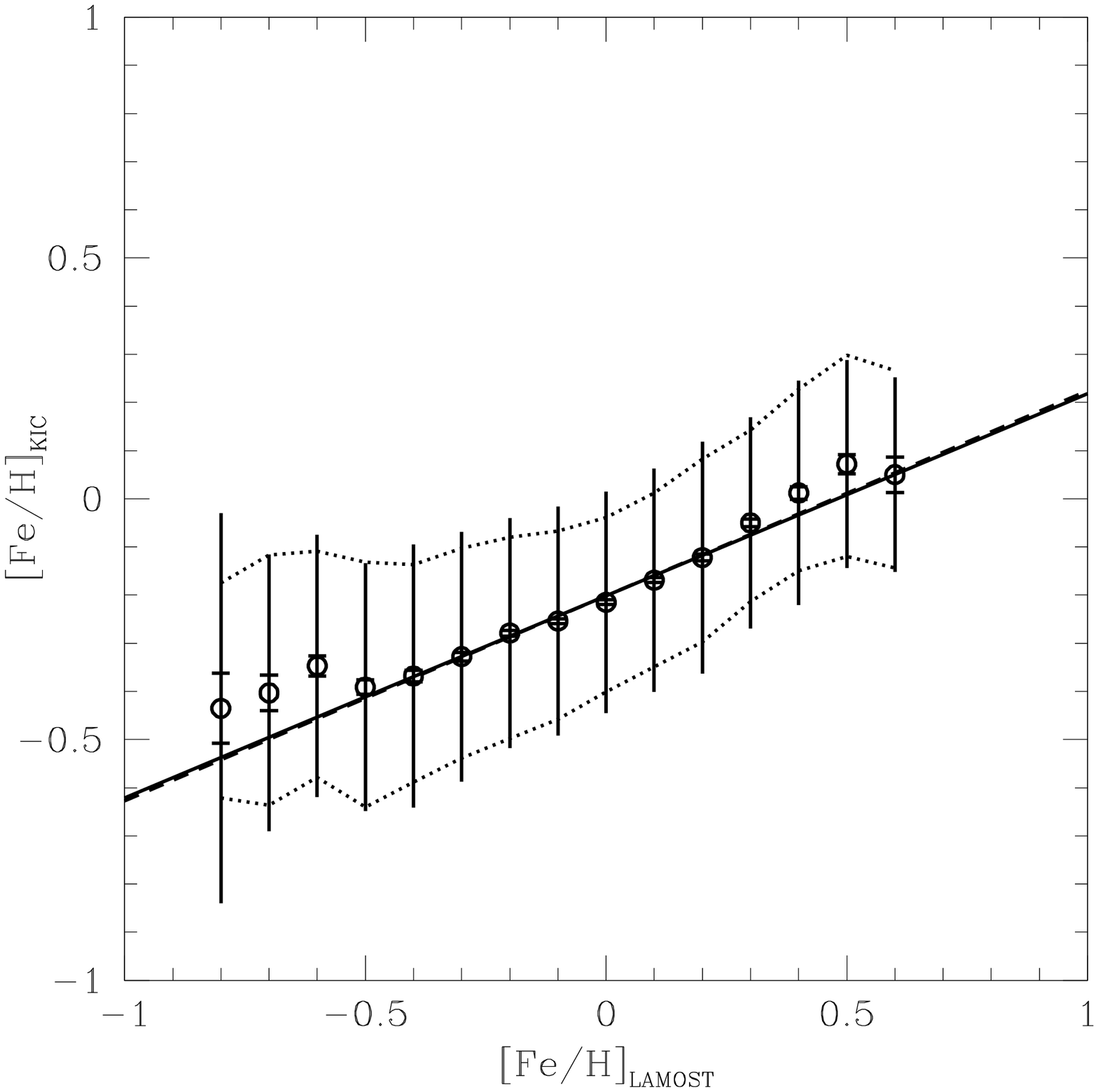}
\caption{\label{fig:bin}
{\it Kepler} star metallicities as determined by KIC as a function of LAMOST 
spectroscopic metallicity. Outer error bars show standard deviations and 
inner error bars show
standard errors of the mean. Dotted lines enclose the central $68.3\%$ 
of the distribution. 
{Solid line is fit to all the data, while
dashed line removes the bins $<200$ stars. The remaining bins
also coincide with the parameter regime where calibrations of LAMOST [Fe/H] with high-resolution spectrostropic [Fe/H] determinations are available ($-0.4 \lesssim {\rm [Fe/H]}_{\rm LAMOST}\lesssim+0.4$). They are essentially the same.  The zero-point and slope
are both detected at high significance, $-0.203\pm 0.002$ and
$0.434\pm 0.011$, respectively.}
}
\end{figure}

Figure \ref{fig:bin} shows a comparison of LAMOST to KIC metallicities
in 0.1 dex bins of LAMOST metallicity.  The outer error bars show the
standard deviation and the inner ones show the standard error of the
mean.  {We makes a linear fit to all the data (the solid line) 
to gain an understanding of the relation between KIC and LAMOST
metallicities to the first order approximation. 
The second and third highest-metallicity bin and the three lowest metallicity bins appear to differ noticeably from the trend. 
The reason for this is unclear. It could be a relatively large 
statistical fluctuation or it could be that either the KIC and/or LAMOST determinations actually change their
trends.  After all, the three highest and lowest metallicity bins only contain $6\%$ of stars in the sample, and stars with ${\rm [Fe/H]}\lesssim-0.4$ or ${\rm [Fe/H]}\gtrsim+0.5$ also belong to the regime of 
the parameter space where we do not have external 
calibrations with high-resolution spectra (see discussions in Section~\ref{sec:high}). To be conservative, we remove these bins, 
each of which has fewer than 200 stars. The dashed line shows the fit to the remaining data. Both the mean offset and slope are detected at very high significance,}
\begin{equation}
{\rm [Fe/H]}_{\rm KIC}
= (-0.203\pm 0.002) + (0.434\pm 0.011){\rm [Fe/H]}_{\rm LAMOST} .
\label{eqn:binfit}
\end{equation}

The scatter in the individual bins is about 0.25 dex. 
We conclude that not only are the KIC metallicities too low,
their dynamic range is substantially compressed relative to the 
metallicity range of the underlying stars. If the above linear relation is inverted to find true metallicity from KIC [Fe/H], 
 the observed scatter is $\sim 0.6\,{\rm dex} = 0.25 \,{\rm dex} /0.43$. 
{We caution that the linear fit given here is to understand 
the systematics of KIC metallicities. Given the large scatter, this 
relation should not be used to ``correct'' the KIC 
metallicity.}

\begin{figure}
\plotone{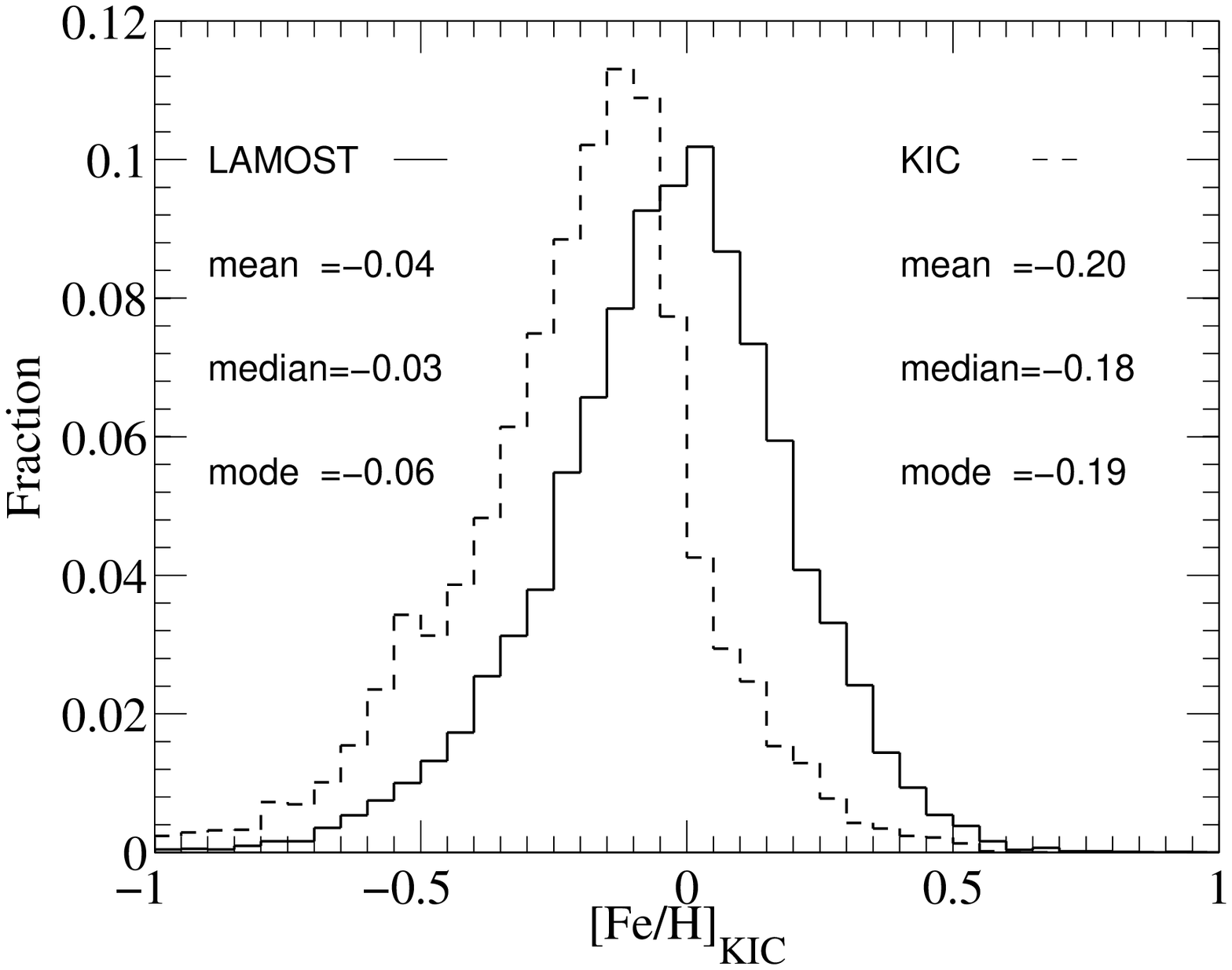}
\caption{\label{fig:met_dist}
Metallicity distribution of the sample. Solid: LAMOST [Fe/H];
Dashed: KIC [Fe/H]. The mean, median and mode of each
distribution are displayed.}
\end{figure}

Figure~\ref{fig:met_dist} shows the metallicity distributions of the
overlapping LAMOST/KIC sample as determined by each catalog. 
Note that the mean LAMOST [Fe/H] is close to solar while the mean KIC [Fe/H]
is about $-0.2$ dex. The LAMOST [Fe/H] distribution is  
similar to that found in the solar neighborhood according to the 
recently revised stellar parameters of the Geneva-Copenhagen Survey \citep{casagrande11}.
\citet{casagrande11}
raised [Fe/H] zero point by about 0.1 dex compared to the previous 
study \citep{nordstrom04}. KIC adopted a Bayesian [Fe/H] prior peaked at 
$-0.1$ dex \citep{KIC}, similar to the distribution from \citet{nordstrom04}.

\section{Comparison of LAMOST to High-resolution Spectroscopic Metallicities 
\label{sec:high}}

\citet{buchhave12} presented the largest homogeneous high-resolution
spectroscopy sample of {\it Kepler} stars. They introduced a new
stellar parameter classification (SPC) technique that reports an
average abundance [M/H] of the elements producing absorption lines
between 5050 \AA\, and 5360 \AA. In order to compare the SPC [M/H] to
LAMOST [Fe/H], we make use of the study by \citet{torres12}, who
systematically examined SPC-determined [M/H] with [Fe/H] as measured
from the widely-used Spectroscopy Made Easy (SME) package
\citep{valenti96} {and the spectral synthesis code MOOG \citep{moog}.}
    
\begin{figure}
\includegraphics[scale=0.7]{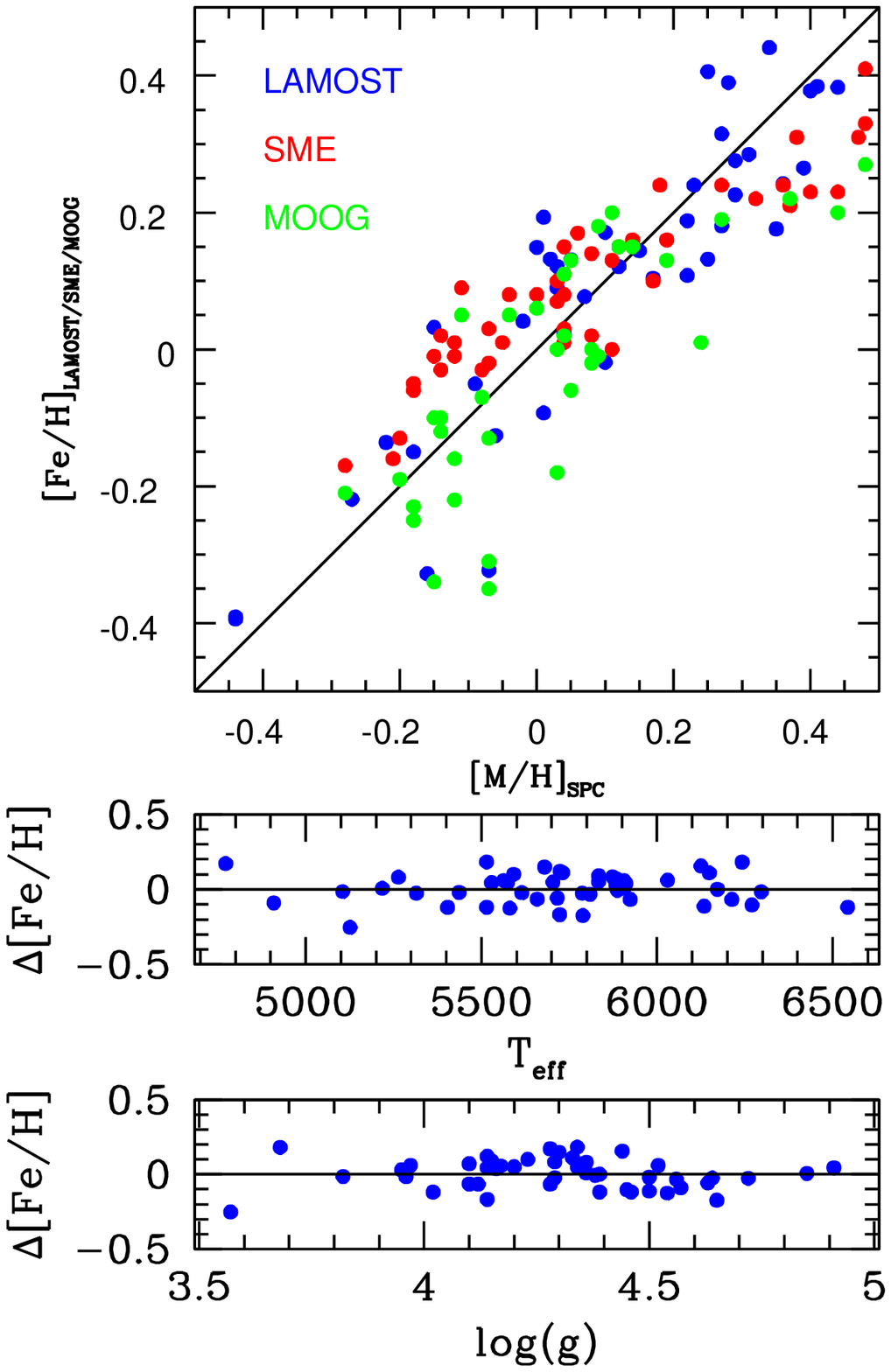}
\caption{\label{fig:comp}
{Comparison of [Fe/H] as determined by LAMOST (blue), the SME technique
(red) or the MOOG technique (green) with [M/H] as determined by the SPC technique \citep{buchhave12}.  The
LAMOST/SPC comparison is based on 47 {\it Kepler} stars in common, while the
SPC/SME and SPC/MOOG comparisons are based on 44 and 36 high-resolution stars, respectively from \citet{torres12}.  
All three comparison show small 
mean offset (LAMOST/SPC: $-0.006 \pm 0.015$ dex, 
SME/SPC: $0.020\pm 0.015$ dex, MOOG/SPC: 
$-0.049\pm0.019$ dex). 
SME/SPC and MOOG/SPC comparisons show trends in different
fashions at amplitudes of $\sim 0.1$ dex, indicating systematics in these methods at this level with unknown sources \citep{torres12}. The standard deviation of the difference between LAMOST and SPC is 
$0.10$ dex, suggesting that LAMOST [Fe/H] determinations are  
reliable at the level that present high-resolution spectroscopic methods are most secure. The middle and lower panels plot the LAMOST/SPC difference as a function of $T_{\rm eff}$ and $\log (g)$, showing no noticeable trends.}
}
\end{figure}

{The upper panel of Figure~\ref{fig:comp} shows the [M/H] by SPC and [Fe/H] by SME of 
44 common stars and by MOOG of 36 common stars} observed with high-resolution spectra from \citet{torres12} in 
filled red circles and green circles, respectively. {They have mean differences of $0.020\pm 0.015$ dex and $-0.049\pm0.019$ dex, 
respectively, but their difference shows 
noticeable trends in difference fashions as a function of metallicity.
These trends have amplitudes at about $0.1$ dex, indicating systematics among these methods at this level, and the sources
of these systematics are unknown \citep{torres12}. The overlap between LAMOST [Fe/H] and SPC [M/H] from 
\citet{buchhave12} is shown in filled blue circles. The 47 common stars
also have a very small mean difference of $-0.006 \pm 0.015$ dex. 
The standard error of the difference is $0.10$ dex, at essentially 
the same level of systematics exhibited in the comparison 
of three different methods. The middle and lower panels of 
Figure~\ref{fig:comp} show the difference between LAMOST [Fe/H] and SPC [M/H] as a function of effective temperature ($T_{\rm eff}$) 
and surface gravity ($\log (g)$), and 
the difference show no noticeable trend over the available $T_{\rm eff}$ and $\log (g)$ ranges.}

The above comparison demonstrates that [Fe/H] measurements from the
LAMOST pipeline are in good agreement with those using
high-resolution spectroscopy over a wide range of metallicity from
$\sim-0.3$\,dex to $\sim+0.4$\,dex.  However it would certainly be
desirable to make more systematical comparisons, especially for
low-metallicity stars. We also note that the overlapping stars between LAMOST 
and \citet{buchhave12} have $5000K\lesssim T_{\rm eff} \lesssim 6500K$, 
which corresponds to the $T_{\rm eff}$ range for the majority of the LAMOST sample.
{The stars used in the \citet{torres12} sample to cross-calibrate SPC, SME and MOOG 
are in the range of $4600K< T_{\rm eff} <6900K$ and $-0.3<{\rm Fe/H}<+0.5$, 
which covers the parameter space of overlapping LAMOST and \citet{buchhave12} stars.
We caution that the reliability of the LAMOST metallicity for stars with 
$T_{\rm eff}$ outside this range shall be examined with other high-resolution spectroscopic data.
Comprehensive calibrations of LAMOST stellar parameters using a 
 large, homogeneous high-resolution spectroscopic sample 
 covering a broader range of parameters are underway.}

\section{Conclusion
\label{sec:conclude}}

In our view, LAMOST metallicities should be used in place of KIC
metallicities whenever they are available (and if there are no
high-resolution spectra available).

And extreme caution is indicated when KIC metallicities are the only
ones available.  In particular, if Equation~(\ref{eqn:binfit}) is inverted
to try to derive real metallicities from KIC metallicities, the observed
scatter in the individual bins (0.25 dex) must be divided by $0.43$
to obtain the final error, i.e., 0.6 dex.

\acknowledgments

We thank Andrew Gould for stimulating discussions. We are grateful to Boaz Katz, Xiaowei Liu, Ali Luo, Yan Wu, and Marc Pinsonneault for helpful comments.
Guoshoujing Telescope (the Large Sky Area Multi-Object Fiber Spectroscopic Telescope LAMOST) is a National Major Scientific Project built by the Chinese Academy of Sciences. Funding for the project has been provided by the National Development and Reform Commission. S.D. is supported by ``the Strategic Priority Research Program-The Emergence of Cosmological Structures'' of the Chinese Academy of Sciences (Grant No. XDB09000000). LAMOST is operated and managed by the National Astronomical Observatories, Chinese Academy of Sciences.
JNF and XHY acknowledge the support from the Joint Fund of Astronomy of National Natural Science Foundation of China (NSFC) and Chinese Academy of Sciences through the Grant U1231202, and the support from the National Basic Research Program of China (973 Program 2014CB845700).

\end{document}